\documentclass[pra,aps,twocolumn,superscriptaddress,showpacs]{revtex4-1}
\usepackage{amssymb}
\usepackage{graphicx}
\usepackage[dvips]{color}
\definecolor{MyOrange}{rgb}{1,0.6471,0}
\begin{document}
\title{Poisson's ratio in cryocrystals under pressure}
\author{Yu. A. Freiman}
%%%%%\email{freiman@ilt.kharkov.ua}
\affiliation{B.Verkin Institute for Low Temperature Physics and
Engineering of the National Academy of Sciences of Ukraine, 47
Lenin avenue, Kharkov, 61103, Ukraine}
\author{Alexei Grechnev}
\affiliation{B.Verkin Institute for Low Temperature Physics and
Engineering of the National Academy of Sciences of Ukraine, 47
Lenin avenue, Kharkov, 61103, Ukraine}
\author{S. M. Tretyak}
\affiliation{B.Verkin Institute for Low Temperature Physics and
Engineering of the National Academy of Sciences of Ukraine, 47
Lenin avenue, Kharkov, 61103, Ukraine}
\author{Alexander F. Goncharov}
\affiliation{Geophysical Laboratory, Carnegie Institution of
Washington, 5251 Broad Branch Road NW, Washington, DC 20015, USA}
\affiliation{Center for Energy Matter in Extreme Environments and
Key Laboratory of Materials Physics, Institute of Solid State
Physics, Chinese Academy of Sciences, 350 Shushanghu Road, Hefei,
Anhui 230031, China}
\author{Eugene Gregoryanz} \affiliation{School of Physics and Centre for Science
at Extreme Conditions, University of Edinburgh, Edinburgh EH9 3JZ,
UK}

\begin{abstract}
{We present results of lattice dynamics calculations of Poisson's
ratio (PR) for solid hydrogen and rare gas solids (He, Ne, Ar, Kr
and Xe) under pressure. Using two complementary approaches - the
semi-empirical many-body calculations and the first-principle
density-functional theory calculations we found three different
types  of pressure dependencies of PR. While  for solid helium PR
monotonically decreases with rising pressure, for Ar, Kr, and Xe
it monotonically increases with pressure. For solid hydrogen and
Ne the pressure dependencies of PR are non-monotonic displaying
rather deep minimums. The role  of the intermolecular potentials
in this diversity of patterns is discussed.}

\end{abstract}
\maketitle

At low temperatures and pressures solid helium is an ultimate
quantum solid displaying such phenomena as zero-temperature
quantum melting and quantum diffusion. As atomic masses and
interatomic forces  increase  in the sequence  Ne, Ar, Kr, and Xe
quantum effects  in their properties become progressively less
pronounced. Solid hydrogen is the only molecular quantum crystal
where both translational and rotational motions of the molecules
are quantum. Translational  quantum effects decrease with
increasing pressure.

Quantum and classical solids respond to the applied pressure
differently. When pressure is applied to a classical solid the
atoms are "pushed into" the hard cores of the potential; as a
result of this core, the compressibility is usually quite small.
Typically, the pressure of 1 GPa results in a few percent change
in molar volume.  At the same time, quantum solids hydrogen and
helium are highly compressible. For hydrogen the pressure of 1 GPa
results in a 100\% change in volume. The physical reason for this
is that the lattice is highly blown up due to the zero-point
kinetic energy. The initial compression works against the weaker
"kinetic pressure" rather than the harder "core pressure".

One of fundamental thermodynamic characteristics describing
behavior of a material under mechanical load  is Poisson's
ratio\cite{Greaves11,Manzhelii97}. For isotropic elastic materials
the Poisson's ratio is uniquely determined  by the ratio of the
bulk modulus $B$ to the shear modulus $G$, which relate to the
change in size and shape respectively\cite{Poirier00}:
\begin{equation}
\sigma=\frac{1}{2}\,\frac{3B/G-2}{3B/G+1}.
\end{equation}
As can be seen from this equation, PR can take values between -1
($B/G\rightarrow0$) and 1/2 ($B/G\rightarrow\infty$). The lower
limit corresponds to the case where the material does not change
its shape and upper limit corresponds to the case when the volume
remains unchanged. Materials with small PR (small $B/G$), such as
cork, are more easily compressed than sheared, whereas those with
PR approaching 1/2 (large $B/G$) are rubber-like: they strongly
resist compression in favor of shear.

For most isotropic materials PR lies in the range $0.2<\sigma<0.5$
\cite{Mott09}. Materials with $0<\sigma<0.2$ are rare - Beryllium
($\sigma=0.03$), Diamond ($\sigma=0.1$) -  and are very
hard\cite{Greaves11}. Typically, PR increases with pressure near
linearly with the rate $\delta\sigma/\delta P\sim 10^{-3}$
(GPa)$^{-1}$ indicating a continuous loss of shear strength
\cite{Steinle99,Zha00,Antonangeli05}.

An unusual pressure dependence of PR decreasing with rising
pressure in solid hydrogen in the pressure range up to 24
GPa\cite{Zha93} and solid helium up to 32 GPa\cite{Zha04} was
found by Zha {\it et al.}. With the aim to investigate the
distinctions in the response of quantum  and classical solids to
the applied pressure we calculated pressure dependencies of PR in
the quantum (He, H$_2$, Ne) and classical (Ar, Kr, Xe)
cryocrystals under pressure. The calculations were performed using
complementary semi-empirical (SE) and density functional theory
(DFT) with generalized gradient approximation (GGA)approaches. The
DFT calculations were performed using the FP-LMTO code RSPt, while
the SE calculations were done using our own code. The calculation
details have been published previously \cite{Freiman14}. It is
important to notice that the two approaches treat solid hydrogen
in fundamentally different ways. SE approach deals with
interaction between H$_2$ molecules, which are treated as nearly
spherically symmetrical quantum rotators, while the DFT can only
treat fully oriented (classical) H$_2$ molecules, ignoring the
zero-point rotations. The $Pca{\rm 2}_{\rm 1}$ oriented structure
has been used for our calculations.

One of the signatures of a quantum crystal is that it melts at
temperature $T_m$ much lower the Debye
temperature\cite{Daniels88}:
\begin{equation}
\Theta_D/T_m \gg 1.
\end{equation}
Figure 1 shows pressure dependencies of $\Theta_D$ and $T_m$ for
solid hydrogen, helium, neon,  and argon. At zero pressure and
temperature the ratio $\Theta_D/T_m$ is infinitely large for
helium and 8.5 for parahydrogen. The ratio rather slowly decreases
with rising pressure. For example, at 1 GPa it is still as high as
3.75 for helium and 3.0 for parahydrogen.  For solid Ne at zero
pressure $\Theta_D/T_m$ = 2.7 thus making solid Ne a candidate for
the manifestation of quantum effects.  Other RGS, with
$\Theta_D/T_m$ = 1; 0.55; and 0.35 for Ar, Kr and Xe respectively,
can be regarded as essentially classical solids.

\begin{figure}
\hspace*{0.6cm}\includegraphics[scale=0.4]{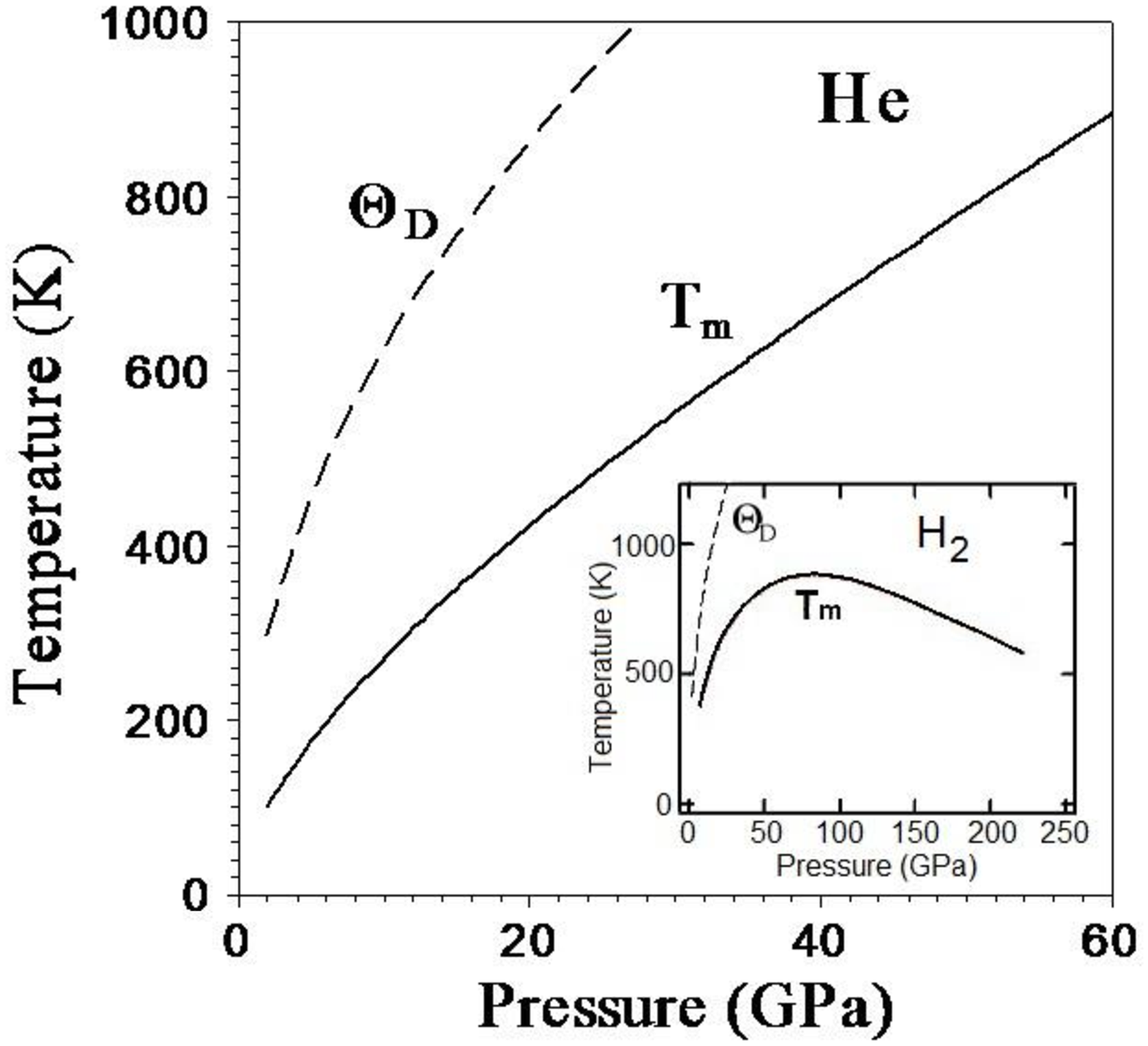}
\includegraphics[scale=0.4]{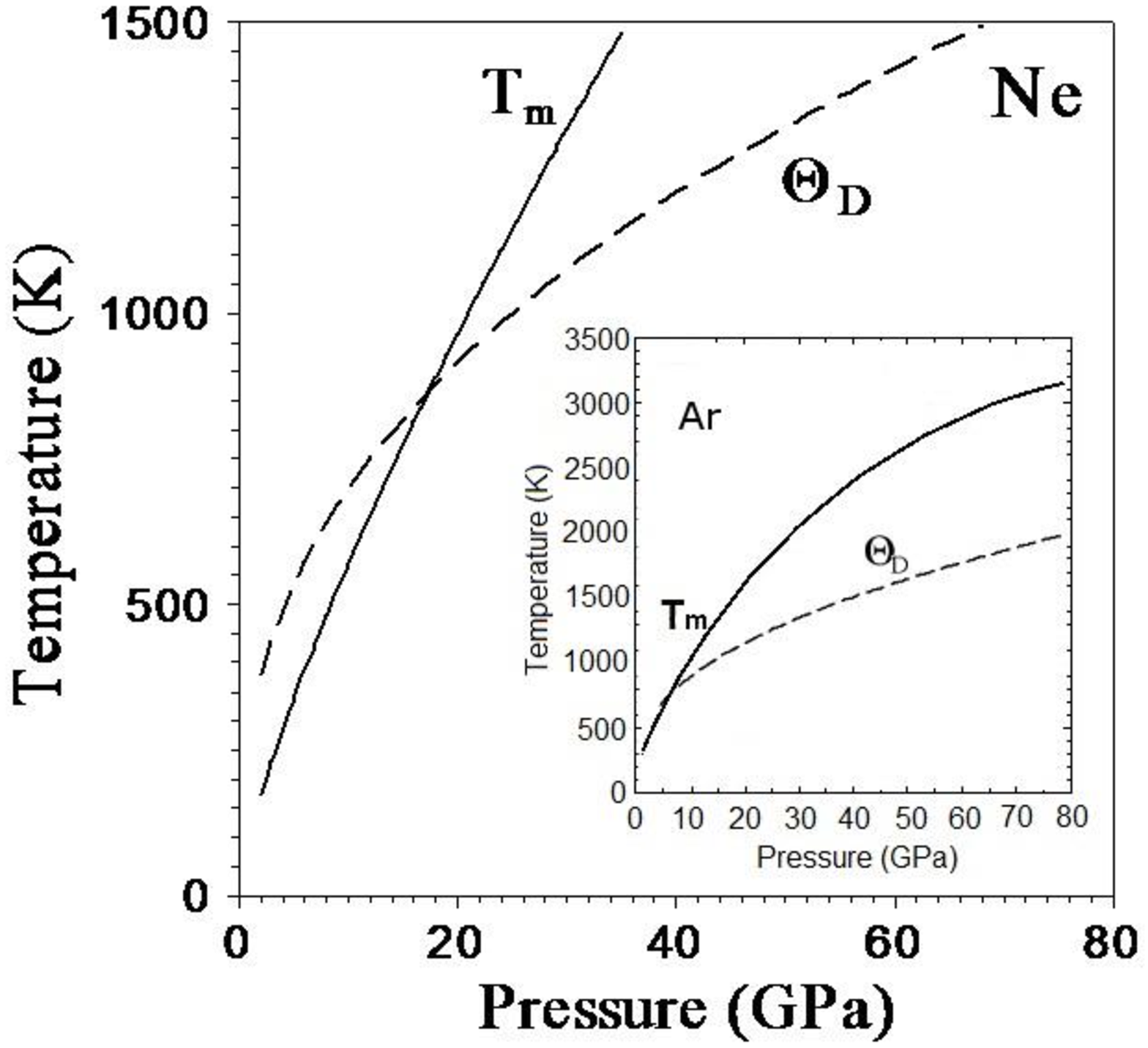}
\caption{\label{f_T_D} Debye temperature  and melting temperature
vs. pressure. Upper Figure:  He and  H$_2$ (insert); bottom
Figure: Ne and Ar (insert). Experimental melting curves: He
 and Ne\cite{Santamaria10}, Ar\cite{Boehler01}. H$_2$ melting curve
 corresponds to Kechin equation \cite{Kechin95}. Debye temperatures
were calculated using the many-body potentials: for He and H$_2$
\cite{Freiman13,Freiman14}, for Ne and Ar \cite{Freiman07}. }
\end{figure}

Rewriting Eq. (1) in terms of the ratio of the bulk (hydrodynamic)
$v_B$ to the transverse (shear) sound velocity $v_S$ we
have\cite{Manzhelii97} :
\begin{equation}
\sigma= \frac{1}{2}\frac{(v_B/v_S)^2-2}{(v_B/v_S)^2-1}.
\end{equation}

The hydrodynamic or bulk sound velocity $v_B$ can be found from
Equation of State:
\begin{equation}
v_B = [\partial P/\partial \rho]^{1/2}= \left[-\,
\frac{V^2}{\mu}\frac{\partial P}{\partial V}\right]^{1/2},
\end{equation}
where $P$ is pressure, $\mu$ is molar mass, and $V$ is molar
volume.  In the calculations  of $v_B$  for H$_2$ we used our SE
and DFT-GGA EOS from Ref. \cite{Freiman12}, for He, Ar and Xe from
Refs. \cite{Freiman07,Freiman08}, and for Kr from Ref.
\cite{Gupta09}. We have also included zero-point vibrations in the
Debye approximation in our calculations of $p(V)$ and $v_B(V)$
\cite{Freiman14}.

Generally, to find sound velocities $v_P$ and $v_S$ one has to
find a complete set of elastic moduli $C_{ij}$. In the case of hcp
lattice there is a simplified scheme based on lattice
dynamics\cite{Metzbower68,Olijnyk00}, which makes it possible to
circumvent the problem of calculations of elastic moduli. In
particular, in this approach it is possible to relate frequency
$\nu$ of the Raman-active $E_{2g}$ phonon mode of hcp lattice and
the shear elastic constant $C_{44}$:
\begin{equation}
C_{44}=\frac{1}{4\sqrt{3}}\frac{c}{a}\frac{m}{a}\nu^2(E_{2g}),
\end{equation}
where $a,\,c$ are the lattice parameters and $m$ is the molecular
mass. The pressure dependencies of $\nu(E_{2g})$ and $C_{44}$ were
found for H$_2$ \cite{Freiman12} and  hcp RGS (hcp He, Ar, Kr, and
Xe) \cite{Freiman08,Shimizu09} using both {\it ab initio} DFT and
semi-empirical  (SE) lattice dynamics approaches. The shear
velocity $v_S$ was obtained using the relation
\begin{equation}
v_S= \sqrt{C_{44}/\rho},
\end{equation}
where $\rho$ is the density, disregarding elastic the anisotropy
of the crystal. A special case is solid Ne which preserves the fcc
structure up to at least 208 GPa\cite{Dewaele08} which makes the
outlined procedure impossible. For this reason for solid Ne we
used results of lattice dynamics calculations by Gupta and
Goyal\cite{Gupta09}.

The sound velocities for H$_2$ and He are given in Ref.
\cite{Freiman13}; the data for Ne by Gupta and Goyal were
published in Ref. \cite{Gupta09}; the data for Ar, Kr, and Xe will
be published elsewhere. Pressure dependencies of Poisson's ratios
for helium, hydrogen, neon, argon, krypton, and xenon calculated
from sound velocities using Eq. (3) are shown in Figs. 2 - 5.

Figure 2 shows the  pressure dependence of Poisson's ratio in
solid He obtained in the framework of SE and DFT-GGA approaches in
comparison with experimental results from
Refs.\cite{Zha04,Nieto12}. Results which account for zero-point
vibrations (ZPV) and those obtained disregarding ZPV are
presented. Both SE and DFT-GGA calculations agree with the
somewhat surprising experimental result of Poisson's ratio
decreasing with pressure. There is a reasonable fair agreement
between the SE theoretical curve (comprising ZPV) and experimental
data. Usually SE results are preferable at smaller pressures while
at higher pressures the DFT approach works better. Comparing the
SE and DFT theoretical curves it is hard to say in which way the
low-pressure SE results could continuously go over to the
high-pressure DFT ones.
%%%{\color{red} [Are you sure you
%%%want to write this??? I don't see any reason to believe there is a
%%%minimum at about 30 GPa, as both SE and GGA curves are
%%%monotonous.] }
It should be noted that the experimental points may show that
around 30 GPa there is a minimum point at the pressure dependence
of PR.

As was said above, typically\cite{Steinle99,Zha00,Antonangeli05}
Poisson's ratio increases with pressure and tends to 1/2 (the
limit of zero compressibility) when pressure goes to infinity. It
would appear reasonable to consider anomalous behavior of PR in
such  quantum solids as He and H$_2$ as a manifestation of quantum
effects. Reasons for such understanding is the following. It is
known that the He and H$_2$ lattices are swelled due to large
zero-point vibrations (ZPV).  If ZPV were not present, "classical"
solid He and H$_2$ would have much smaller zero molar volumes
($V_0^{\rm cl}({\rm He})\approx$11.2 cm$^3$/mol; $V_0^{\rm
cl}({\rm H_2})\approx$7.4 cm$^3$/mol), i.e. the swelling effect is
huge \cite{Freiman08}. Until the volume reaches about $V_0^{\rm
cl}$, the main effect of the external pressure is the suppression
of the zero-point vibrations and not the compression of the
electron shells.

\begin{figure}
\vspace*{0.6cm}
\includegraphics[scale=0.29]{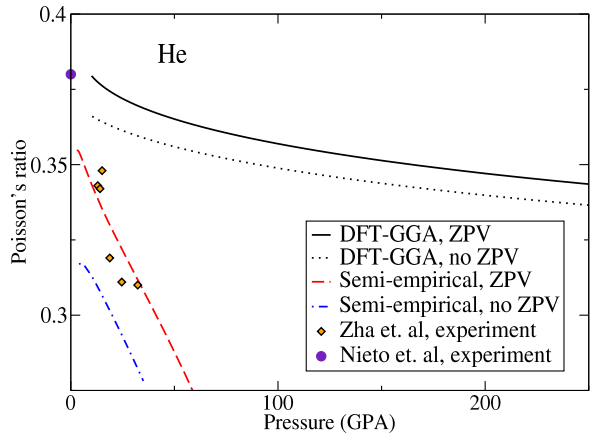}
\caption{\label{f_he}(Color online). Poisson's ratio of solid He
as a function of pressure. Theory: this work, Experiment: Zha {\it
et al.}\cite{Zha04}; Nieto {\it et al.}\cite{Nieto12}.}
\end{figure}
\begin{figure}
\vspace*{0.65cm}
\includegraphics[scale=0.285]{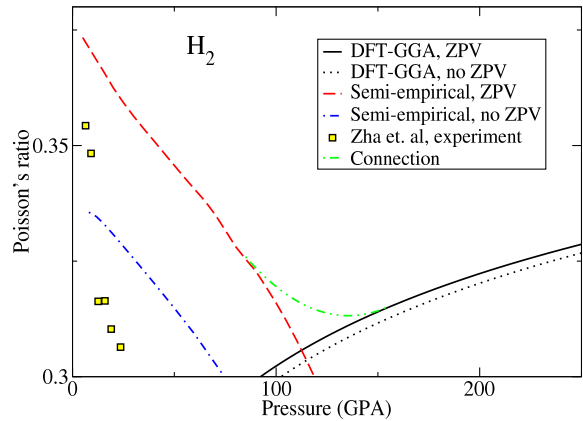}
\caption{\label{f_h2} (Color online). Poisson's ratio  of solid
H$_2$ as a function of pressure. Theory: this work, Experiment:
Zha {\it et al.}\cite{Zha93}}
\end{figure}

To check whether this explanation is correct we calculated PR of
He disregarding ZPV, that is,  for "classical" He both  in the SE
and DFT approaches (dot-dash and dotted curves, respectively,
Fig.2). As can be seen, the pressure dependence  of PR with and
without ZPV is qualitatively the same. Thus, the anomalous
(descending with rising pressure) behavior of PR is not a quantum
effect. As can be seen from Fig. 2, the contribution of ZPV into
PR is positive. This fact is easily understood if we take into
account that the introduction of ZPV is a step to liquation but PR
of  liquid is an upper bound for PR of any substance. Naturally,
the relative value of this contribution increases with decreasing
pressure and as pressure goes to zero it increases up to 15\%. The
effect of ZPV is much higher in the case of $^3$He. Nieto {\it et
al.}\cite{Nieto12} showed that the mixture $^3$He-$^4$He has
higher PR than pure $^4$He. For pure $^3$He they gave value of PR
0.473 rather close to the liquid limit.

 \begin{figure}
\includegraphics[scale=0.5]{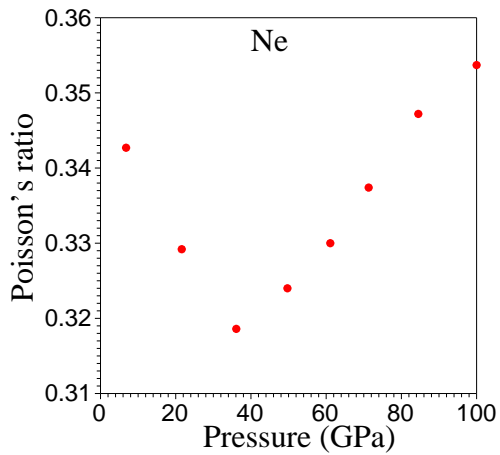}
\caption{\label{f_ne} (Color online) Poisson's ratio of solid Ne
as a function of pressure. Calculated using data on sound
velocities by Gupta and Goyal\cite{Gupta09}.}
\end{figure}

\begin{figure}
\includegraphics[scale=0.5]{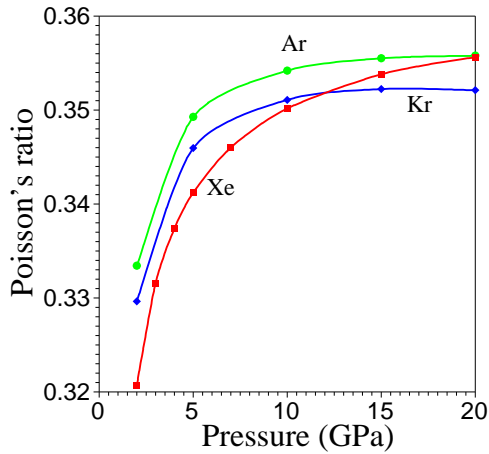}
\caption{\label{f_ar} (Color online) Poisson's ratio of solid Ar,
Kr, Xe as a function of pressure. Calculated using SE data on
sound velocities (unpublished).}
\end{figure}

The theoretical and experimental pressure dependencies of PR for
solid H$_2$ are shown in Fig. 3. As can be seen, the SE and
DFT-GGA approaches give the opposite signs of the pressure effect
on PR: PR decreases with rising pressure for SE and increases for
DFT-GGA. Since in the experimentally studied pressure range (up to
24 GPa) the SE result agrees qualitatively with experiment
\cite{Zha93},  we conclude that at low pressures PR decreases with
rising pressure for solid H$_2$.  It is known that while the SE
approach works well for molecular solids at low pressures, for
higher pressures the DFT -GGA approach is preferable. Thus the
PR$(p)$ curve can be subdivided into three regions: At the
low-pressure region SE is expected to work well, while at high
pressures  we can use the DFT-GGA approach. In the intermediate
pressure range both approaches fail. The dot-dot dash curve shows
schematically a possible continuous transition from the
low-pressure asymptote to high-pressure one. Resulting pressure
dependence  of PR for H$_2$ is non-monotonic displaying rather
deep minimum. It should be noted that the transient region from
the descending to the ascending curves falls on phase II of the
hydrogen phase diagram.  As mentioned above, the SE and DFT
approaches treat the orientational degrees of freedom in H$_2$ in
completely different ways: the former regards H$_2$ molecules as
nearly spherically symmetric quantum rotators (as in phase I),
while the latter considers classically oriented H$_2$ molecules
(as in phase III), completely ignoring any quantum rotations or
librations. It seems likely that this is the reason why SE and DFT
give such drastically different PR$(p)$ curves for H$_2$, while
results for helium are qualitatively similar. It would mean that
the PR minimum in hydrogen is related to the orientational
transition at around $110$ GPa, however more detailed study of
this question is beyond the scope of the present work.

A similar curve  with a deep minimum was obtained for PR in solid
neon (Fig. 4).  The pressure dependence of PR was obtained from
the SE theoretical results on sound velocities obtained by Gupta
and  Goyal\cite{Gupta09}. Unfortunately, experimental data on
sound velocities in solid Ne exist for very narrow pressure range
5 - 7 GPa\cite{Shimizu05}. In this region PR $\approx$ 0.37.

Figure 5 shows the pressure dependencies of PR obtained in the SE
approach for Ar, Kr, and Xe. In contrast with He, H$_2$, and Ne,
we obtained that PR for the heavy RGS increases with rising
pressure.

In conclusion, we present results of lattice dynamics calculations
of Poisson's ratio for solid hydrogen and rare gas solids (He, Ne,
Ar, Kr and Xe) under pressure. Using two complementary approaches:
lattice dynamics  based on the semi-empirical many-body potentials
and {\it ab initio} DFT-GGA we found three different types of the
behavior of PR with pressure. While for solid He PR monotonically
decreases with rising pressure, for Ar, Kr, and Xe it
monotonically increases with pressure. For solid H$_2$  and Ne PR
are non-monotonic with pressure displaying rather deep minimums.
To investigate the role of quantum effects we performed the
calculations of PR disregarding zero-point vibrations and found
qualitatively similar results, that is, we proved that the effects
have a non-quantum origin. or{blue}  We may rather say that the
anomalies, discovered for H$_2$, He and Ne,  and quantum effects
in these cryocrystals have common origins: weak intermolecular
interactions and small masses of constituent atoms and molecules.

We thank  J. Peter Toennies  for valuable discussions.

\end{document}